\documentstyle[aps,twocolumn,epsf,amssymb,prl]{revtex}
\begin{document}
\bibliographystyle{prsty} 
\title{Competition between zero bias anomaly and proximity effect in
disordered systems}

\author{Yuval Oreg$^a$, P.~W.~Brouwer$^a$, B.~D.~Simons$^b$, and Alexander
   Altland$^c$} 
\address{ 
$^a$ Lyman Laboratory of Physics, Harvard
   University, Cambridge, MA 02138, USA \\ 
$^b$ Cavendish Laboratory,
   Madingley Road, Cambridge, CB3 0HE, United Kingdom \\ 
$^c$ Theoretische Physik III,
Ruhr-Universit\"at Bochum, 44780 Bochum, Germany\\
  {}~{ }
  \medskip \\
  \parbox{14cm} {\rm
  We investigate the suppression of the proximity effect in mesoscopic
  normal metal/superconductor systems induced by Coulomb interaction.
  We identify and elucidate the mechanism by which disorder leads to
  an amplification of this effect. In particular, for strong enough
  disorder, the proximity effect is shown to vanish.  An expression
  for the tunneling density of states is obtained, and experimental
  applications are discussed.
\vspace{-0.3cm}}}
\maketitle
The influence of a superconductor (S) brought into contact with a
normal (N) disordered metal has drawn great interest recently both
theoretically~\cite{NS:Bauernschmitt94,NS:Golubov95,NS:Beenakker95,NS:Melsen96,NS:Belzig96,NS:Shechter97,NS:Zhou98}
and experimentally~\cite{NS:Gueron96}.  In such a geometry, the
superconductor tends to impart aspects of its superconducting
character on to the normal region. Underlying this effect is the
mechanism of Andreev scattering by which an electron from the normal
region is reflected at the SN-interface as a hole, and a Cooper pair
is added to the superconducting condensate.  This phenomenon has a
striking effect on spectral and transport properties, known
collectively as the proximity effect (PE)~\cite{NS:Beenakker95}. In
particular, if the normal region is finite, a minigap $E_{\rm g}$
develops in the excitation spectrum, the size of which depends on the
typical time between successive Andreev reflections for quasiparticles
in the normal metal, which in turn depends on the transparency ${\cal
T}$ of the SN-interface~\cite{NS:Melsen96,NS:Belzig96}.
 
The minigap in the excitation spectrum can be measured through the
tunneling conductance of an external probe weakly coupled to the
normal system. In the absence of a superconductor, the repulsive
electron-electron ($e$--$e$) interactions in the normal metal give
rise to a suppression of the tunneling density of states (TDoS) at
small voltages~\cite{DS:Altshuler79,DS:Finkelstein90,DS:Levitov96}.
This phenomenon, known as the zero bias anomaly (ZBA), can be
understood as the partial blocking of electron tunneling due to
Coulomb repulsion from its own charge cloud~\cite{DS:Levitov96}.
Since in low-dimensional dirty metals the electron charge spreads more
slowly than in clean metals, this effect is enhanced.  One therefore
expects that it will play an important role in small dirty SN
structures, where the ZBA is combined with the gap induced by the PE.

 Although various important aspects of the role of Coulomb
interactions in mesoscopic SN (and S) structures have already been
investigated \cite{NS:Zhou98,DS:Finkelstein90,NS:Huck97,NS:Nazarov96},
the disorder amplified interplay between `Coulomb blocking' and PE
described above has so far not been discussed.

To be specific, it will be shown in this Letter how the mechanism of
Coulomb blocking renormalizes the transparency of the tunnel barrier
at the SN interface, thereby leading to a suppression of the minigap
$E_{\rm g}$. In particular, we find that, above a critical strength of
disorder, the minigap vanishes altogether.  In addition to affecting
the size of the gap, the same mechanism also leads to a (non-singular)
suppression of the TDoS. (Note that, unlike in the absence of Coulomb
interaction, the position of the gap and the enhancement of the TDoS
are not constrained by sum-rules.)  Although in this Letter the focus
will be on the behavior of the TDoS, it is worth mentioning that other
properties (e.g. the Josephson current) are also affected by the
suppression mechanism above.

Such behavior has important experimental consequences: If the normal
metal is relatively clean, the Coulomb blocking effect is weak. At
temperatures $T\ll E_{\rm g}$, a measurement of the differential
conductance $dI/dV$ as a function of voltage $V=\epsilon/e$ for a
current injected from an external probe into the normal part of the SN
structure reveals an enhancement of the TDoS above a sharp minigap
edge.  As the effective \cite{NS:EffectiveDisorder98} strength of the
disorder in the normal region is increased both the minigap $E_{\rm
g}$ and the enhancement near $E_{\rm g}$ are diminished and a
superimposed signature of the standard ZBA suppression in the TDoS
appears. At a critical value of disorder the PE disappears altogether,
and only the conventional ZBA remains.

 To quantitatively study the PE in an interacting environment, we
employ the non-linear $\sigma$-model for disordered
superconductors~\cite{DSC:Kravtsov91,NS:Altland98}. The conventional
quasi-classical picture~\cite{NS:Eilenberger68,NS:Usadel70} of the PE
is obtained by evaluating the (non-interacting) $\sigma$-model
functional in a saddle-point (SP) approximation\cite{NS:Altland98}. We
then combine this method with the standard replica field theoretic
approach to study Coulomb interaction effects in disordered
conductors~\cite{DS:Finkelstein90}. Following this strategy, rather
than conventional diagrammatic perturbation theory, is motivated by
the fact that the latter is inapplicable in PE dominated environments
(see below).
 
Although the theory is not restricted to a particular geometry, to be
concrete we consider a diffusive normal metallic film of width $d$
separated from a clean bulk superconductor by a tunnel
barrier with transparency ${\cal T}$. In the dirty limit,
$(\varepsilon,\Delta)<\tau^{-1}\ll E_F$, the impurity averaged (replicated)
quantum partition function is obtained from a functional field
integral involving the low energy effective
action~\cite{NS:Altland98,NS:footnote98}
 
\begin{equation}
\begin{array}{ll}
 {S[Q]}=\frac{\pi \nu}{8}\int d{\bf r}\;{\rm tr}\big\{D (\nabla Q)^2
 -4 (V+\Sigma) Q-\\
 \qquad\qquad\vphantom{\bigg|}-2\gamma\delta(z)\left[Q(z^+)-Q(z^-)\right]^2\big\}+\\
 \qquad\qquad+{1\over 4 T}\int d{\bf r} \; d{\bf r}'\; {\rm
 tr}\left[V({\bf r}) \Gamma_0^{-1}({\bf r}-{\bf r}')V({\bf r}')
 \right],
\end{array}
 \label{eq:Saction}
\end{equation}
where $D$ denotes the bare diffusion constant, $\Sigma({\bf r})= (
\Delta({\bf r}),0,\hat{\varepsilon}) \cdot {\bbox{\sigma}}^{\rm ph}$,
$\hat{\varepsilon}$ represents the diagonal matrix of Fermion
Matsubara frequencies $\varepsilon_n=\pi T(2n+1)$, $\tau$ is the mean
free time, $E_F$ is the Fermi energy, and the order parameter
$\Delta$, specified in an appropriate gauge, takes a constant non-zero
value in the superconductor and vanishes in the normal region.  In our
slab geometry the film is in the $xy$ plane the interface is located at
$z=0$ and the metal at $z<0$.  The functional integration extends over
matrix fields $Q=\hat T \left(\Lambda\otimes\sigma_3^{\rm ph} \right)
\hat T^{-1}$, where ${\bf \sigma}^{\rm ph}$ denote Pauli matrices
operating in the particle/hole sector, $\Lambda={\rm
sgn}(\hat{\varepsilon})$, and the generators $\hat T$ have degrees of
freedom corresponding to the spin and particle/hole sectors, as well
as replica indices and Matsubara frequencies \cite{NS:footnote98}.
The strength of coupling across the SN-interface is controlled by
$\gamma=cv_F{\cal T}/(1-{\cal T})$, where ${\cal T}$ denotes the
transparency of the tunnel barrier and $c$ represents a numerical
constant of order unity~\cite{NS:Kuprianov88}. In the following we
assume that $\gamma/d \ll E_d \ll \Delta$, where $E_d = D/d^2$ is the
inverse transport time across the normal region. As charge transport
across the SN-interface is mediated by the coupling term $\propto
\gamma$ in action (\ref{eq:Saction}), the fields are subject to the
boundary condition on the normal derivative $\partial_z Q|_{z=0}=0$ at
the SN-interface.  Finally, the influence of the Coulomb interaction
enters the effective action through the Hubbard-Stratonovich field
$V$. Here attention is limited to the density-density channel which
gives rise to singular modifications of the SN-coupling. The repulsive
interaction in the Cooper channel suppresses the minigap however not
in a singular manner\cite{NS:Zhou98}.  In deriving the low energy
effective action, several mechanisms of renormalization have already
been taken into account: Firstly, the repulsive $e$--$e$ interaction
in the bulk superconductor leads to high energy renormalization of the
implicit electron-phonon coupling, which in turn modifies the
superconducting order parameter~\cite{SC:Morel62}. Secondly, the
kernel $\Gamma_0^{-1}$ includes screening effects both from standard
random phase approximation corrections~\cite{DS:Finkelstein90}, as
well as from nearby metallic (or superconducting) objects
\cite{NS:Phase98}.
 
Due to the presence of the order parameter $\Delta$, standard
perturbative approaches\cite{DS:Finkelstein90} for evaluating the
action (\ref{eq:Saction}) fail.  Following Ref.~\cite{NS:Altland98},
we therefore begin by subjecting the action to a mean-field
analysis. A variation of the action~(\ref{eq:Saction}) with respect to
$Q$ subject to the constraint $Q^2=\openone$, generates the SP
equation
\begin{eqnarray}
\label{eq:saddle}
D \nabla (Q \nabla Q) + [Q, \Sigma+V] &=& \gamma [Q(0^{+})
\delta(z-0^{-}),Q] \nonumber \\ && \mbox{} + \gamma[Q(0^{-})
\delta(z-0^{+}),Q].
\end{eqnarray}
In the non-interacting theory, this equation is the conventional
Usadel equation~\cite{NS:Usadel70} describing the spatial dependence
of the average quasi-classical Green function. The right hand side of
Eq.~(\ref{eq:saddle}), together with $\partial_z Q|_{z=0}$ reproduces
the boundary conditions to the Usadel equation in the presence of a
tunneling barrier between N and S \cite{NS:Kuprianov88}.  These
boundary conditions implies that in the low energy regime, $\varepsilon
\ll E_d$, $Q$ does not fluctuate across the N layer. (The minimum
energetic cost associated with a fluctuating mode compatible with the
boundary conditions is of ${\cal O}(E_d)$.)  As a consequence the
boundary $\delta$-function $\delta(z-0^+)$ can be replaced by $
d^{-1}$ which substantially simplifies the solution of
(\ref{eq:saddle}).

We begin by investigating the non-interacting system: $V=0$. It is 
straightforward to check that in the low energy regime Eq.
(\ref{eq:saddle}) is solved by $Q(z < 0)\simeq\sigma^{\rm ph}_1
\otimes\openone$ \cite{NS:Phase98} and,
\begin{equation}
\label{eq:NIQ}
Q(z>0)\equiv \Lambda_\gamma \equiv \cos \hat{\theta}\otimes\sigma^{\rm ph}_3 +
\sin\hat{\theta} \otimes\sigma^{\rm ph}_1,
\end{equation}
where the diagonal matrix $\hat{\theta}$ has elements $\theta_n=
\arccos(\varepsilon_n/E_n)$ with $E_n^2=\varepsilon_n^2+(E_{\rm
  g}^{0})^2$ and $E_{\rm g}^0=\gamma/d$. This result implies the
existence of a minigap in the excitation spectrum of the normal region
of size $E_{\rm g}^0 = \gamma/ d \ll E_d$. Applied to the TDoS,
$\nu=\nu_0{\rm Re}\langle {\rm tr} ({\cal P}^{R}\otimes \openone_s
\otimes \sigma_3^{\rm ph} \otimes\Lambda ) Q \rangle_Q/4$, where
$\nu_0$ denotes the DoS of the bulk normal metal, $\openone_s$ is the
identity matrix in the spin sector and ${\cal P}^R$ a projector in
replica space\cite{DS:Finkelstein90}, the mean-field solution implies
a square root singularity at $E_{\rm g}^0$.  (In passing we note that,
due to the smallness of the SN-coupling, invoking a self-consistent
renormalization scheme for the order parameter near the interface is
unnecessary~\cite{NS:Likharev79}.)

Restoring the Coulomb interaction field $V$, the SP
equation~(\ref{eq:saddle}) becomes non-diagonal in Matsubara space. An
exact solution being now unavailable, a natural approach would be to
seek a perturbative solution of the SP equation. However, as we will
see below, such an analysis fails to properly account for the
renormalization of the coupling constant $\gamma$. While the former
leads to a suppression of the TDoS above $E_{\rm g}^0$, the latter
leads to an explicit renormalization of the gap itself. This effect
parallels the renormalization of $E_{\rm g}^0$ encountered in
non-interacting disordered SN-structures due to mechanisms of weak
localization (in the particle/hole channel)~\cite{NS:Altland98}. Our
approach is to include both effects systematically.
 
Beginning with a perturbative analysis of the SP equation
~\cite{DSC:Kamenev98remark}, we seek a transformation in the normal
region,
\begin{equation}
\label{eq:param}
Q(0<z<d)=e^{-W/2} \Lambda_\gamma e^{W/2},\ \ \
\{W,\Lambda_\gamma \}=0,
\end{equation}
which satisfies Eq.~(\ref{eq:saddle}) to leading order in both the
generators $W$ and $V$. Substituting this ansatz into the SP equation
and linearizing we obtain

\begin{eqnarray}
D\partial^2 W - \{W,\hat E\} = [V,\Lambda_\gamma],
\end{eqnarray}
where the diagonal matrix $\hat E = \{E_n\}$. 
For the slab geometry, the solution is given by
\begin{eqnarray}
\label{eq:D}
W_{nm}(q)&=&  \frac{2V_{nm}}{E_m+E_n+Dq^2} 
\left[\left(\cos\theta_n-\cos \theta_m \right)\sigma_3 \right.+
\nonumber \\ && \mbox{} \left. +
\left(\sin\theta_n-\sin \theta_m \right) \sigma_1  \right] \otimes\openone,
\end{eqnarray}
where $q$ is the momentum in the $xy$-plane and the denominator
has the significance of a generalized diffusion pole (which, in the
metallic limit $E_{\rm g}\to 0$, approaches the standard form of a
diffuson.)

Substituting Eq.~(\ref{eq:param}) into the action we obtain an
effective action for the field $V$, describing how the screening
properties of the normal metal are modified as a result of the
diffusive motion in the presence of the superconductor.  In
particular, the $e$--$e$ interaction kernel $\Gamma_0^{-1}\to
\Gamma_0^{-1}+\hat \Pi\equiv \Gamma^{-1}$, where the operator $\hat
\Pi$ accounts for the anomalous polarizability of the PE influenced
metal.  The dominant contribution to the TDoS derives from processes
where only small momenta $q \ll 1/d$ and energies $\omega \ll E_d$ are
transferred by the Coulomb interaction.  In this case screening due to
both the superconductor and the polarization operator $\hat \Pi$
render the Coulomb interaction effectively short range,
$\Gamma^{-1}({\bf r}-{\bf r}')\approx \lambda^{-1}$, where $\lambda$
is a constant. (We remark that a more precise treatment of the
screening properties may be important in other geometries.)
 
Using the Coulomb corrected SP solution above to compute the zero
temperature TDoS, after summation over Matsubara frequencies and
analytic continuation $\varepsilon_n \to i\varepsilon+ 0^+$ we obtain
the correction \cite{NS:highenergy98}
\begin{equation}
 \label{eq:nuI}
\begin{array}{l}
  {\delta \nu(\varepsilon)\over \nu_0}= -\frac{1}{4}\lambda t
 \varepsilon(\varepsilon^2-\left. E^0_{\rm g} \right.^2)^{-1/2} \\
 \qquad\qquad \vphantom{\bigg|}\mbox{} \times \left[ \log{2 E_d\over
 E^0_{\rm g}} + f(\varepsilon/E^0_{\rm g}) \right] \theta(\varepsilon
 - E^0_{\rm g}),\\ f(x) = {2 - x^2 \over x} (x^2-1)^{-1/2} \log
 \left[x + (x^2-1)^{1/2} \right],
\end{array}
\end{equation}
where $\rho$ is the resistivity of the thin normal layer, $t=(e^2/2
\pi^2 \hbar) (\rho/d)$ controls the strength of disorder and
$\theta(x) = 1$ for $x > 0$ and $0$ otherwise.  Noting that $f(x) \sim
- \log 2x $ for $x \gg 1$ and $f(1)=1$, we see that Eq.~(\ref{eq:nuI})
reproduces the usual ZBA for $\varepsilon \gg E^0_{\rm g}$ and
saturates to a finite value at energies close to $E^0_{\rm
g}$. Significantly, no states appear below $E_{\rm g}^{0}$. From
Eq.~(\ref{eq:nuI}) we see that the strength of the square root
singularity is suppressed by a logarithmic contribution, the
coefficient inside the $\log$ being slightly larger than the
coefficient in the usual ZBA. There are two effects that distinguish
this contribution from the ZBA in the absence of the superconductor:
Firstly, due to the enhancement of the TDoS in the vicinity of
$E^0_{\rm g}$, the single electron propagator decays (in time) slower
than in a normal metal, thereby enhancing the Coulomb blocking
effect. Secondly, the conductance near $E_{\rm g}^{0}$ is larger so
that the cloud can relax faster, which leads to a smaller interaction
correction. As can be seen from Eq.~(\ref{eq:nuI}), the combination of
these effects does not lead to singular corrections to the TDoS.
 
Eq.~(\ref{eq:nuI}) predicts that the position $E_{\rm g}$ of the TDoS
gap is unaffected by interactions, thereby indicating that our so-far
perturbative analysis is qualitatively incomplete: We expect that
Coulomb blocking leads to a downward renormalization of the coupling
strength between N and S, and hence to a shrinkage of the gap itself.
To quantitatively describe this effect, we follow
Ref.~\onlinecite{DS:Finkelstein90}, and employ a renormalization group
(RG) procedure in which the fast, high energy degrees of freedom
renormalize the slow.
 
Eq.~(\ref{eq:nuI}) shows that the main (logarithmically divergent)
contribution to the TDoS stems from energies larger than the
(renormalized) minigap $E_{\rm g}$ where the system behaves
effectively like a metallic one. This observation dictates the
strategy of the RG approach: Employing the (renormalized) $E_{\rm g}$
itself as an infrared RG cutoff we assume that for energies
$\varepsilon > E_{\rm g} $, the system can be approximated by a
conventional 2D disordered conductor. For energies $\varepsilon\gg
E_{\rm g}$ the treatment is accurate, while for $\varepsilon\sim
E_{\rm g}$ we expect corrections which are small in the limit $E_d\gg
E_{\rm g}$ \cite{NS:inlog98}. In the ultraviolet, the RG procedure is
cut off by the ``high'' energy scale $E_d$.

The separation of energy scales above allows us to apply, without
technical modifications, the standard RG procedure for 2D disordered
systems~\cite{DS:Finkelstein90} in the interval $E_{\rm
g}<(Dq^2,\varepsilon)<E_d$. Conceptually, the only difference is that
the perturbative RG is superimposed on a background of the Usadel
solution. Carrying out the RG procedure, yields a renormalization of
the coupling strength $\gamma$, \cite{DS:Finkelstein90,DS:Levitov96}
\begin{equation}
\label{eq:gamma}
\gamma (\varepsilon) = \gamma_0 e^{-(t \lambda/4)  \log(E_d /\varepsilon)}.
\end{equation}
Combined with the relation $E_{\rm g} = \gamma/d$ we obtain a
self-consistent equation for $E_{\rm g}$,
\begin{equation}
\label{eq:selfconst}
E_{\rm g} = E_{\rm g}^0 e^{- \alpha \log E_d/E_{\rm g}},
\end{equation}
where $\alpha = t\lambda/4$. The solution of Eq.(\ref{eq:selfconst})
represents the renormalized energy gap in the normal metal layer,
\begin{equation}
\label{eq:selfconstsol}
\textstyle E^*_{\rm g} = E_{\rm g}^0 \left( \frac{ E^0_{\rm g}} {E_d}
\right)^{\frac{ \alpha}{1-\alpha}}, \;\;\ \alpha < 1.
\end{equation}
Hence, the gap decreases monotonically with increasing $\alpha$ (or
disorder $t$) up to a critical value $\alpha=1$ above which $E_{\rm
g}^{*}$ vanishes altogether.
 
Treating the tunneling from the probe on the same footing (i.e.
renormalization above $E_{\rm g}$ as in the case of a normal metal,
and applying a cut-off at $E_{\rm g}$) leads to an exponentiation of
the first order correction to the
TDoS~\cite{DS:Finkelstein90,DS:Levitov96}.  Thus, the asymptotic
regimes above and below $E^*_{\rm g}$ can be summarized as
 \[
\begin{array}{l}
 {\nu(\varepsilon)\over \nu_0}=\left\{\begin{array}{ll}
\varepsilon e^{-\alpha \log (E_d/
 \varepsilon)} \left[\varepsilon^2-E_{\rm g}^2(\varepsilon)\right]^{-1/2}
 &\varepsilon \gg E_{\rm g}^*\\
 0 &\varepsilon <E_{\rm
 g}^*
\end{array}\right.\\
 E_{\rm g}(\varepsilon)=E_{\rm g}^0 e^{- \alpha t\log(E_d/ \varepsilon)},
 \quad E_{\rm g}(\varepsilon=E^*_{\rm g})= E^*_{\rm g}.
\end{array}
\]

Before concluding, a few additional remarks are in order. First, in the RG
treatment employed here, the exchange and Cooper channels, as well as
the renormalization of the diffusion constant and frequency by
interaction and weak localization effects~\cite{DS:Finkelstein90}
where neglected. Since there is no reason to expect that
these terms are small in the strong disorder case when
$\log[E_d/\varepsilon ]>1/t$, the accuracy of the present theory is
limited. Since we expect all the terms we neglected
to further suppress the PE, we believe that their effect will be
to decrease the critical strength of disorder below $\alpha =1$.

Secondly, it is instructive to interpret the Coulomb renormalization
of the proximity effect discussed above from an alternative point of
view.  The fields $Q$ can be subjected to a quantum mechanical gauge
transformation $Q\to e^{i\Phi} Q e^{-i\Phi}$. An interesting
reformulation of the theory arises if $\Phi$ is chosen so as to
eliminate the most singular momentum/frequency components of the
disorder enhanced Coulomb potential $V$. As for the case of extended
N-systems, it has recently been shown\cite{DSC:Kamenev98,DS:Kopietz98}
that gauging the theory in this way, leads to the nonperturbative
expression for the ZBA \cite{DS:Finkelstein90}. In the problem
considered here, the situation is different, the gauge transformation
not only removes the singular components of the interaction but also
directly couples to the phase of the order parameter in the
superconductor. More specifically, gauging the theory as outlined
above, induces dynamical boundary phase fluctuations. These
effectively reduce the SN-coupling and hence suppress the proximity
effect. These considerations indicate that the gauge
transformation\cite{DSC:Kamenev98,DS:Kopietz98} effectively mediates
between the phase and the charge representation\cite{NS:Schon90} of
charging phenomena in disordered SN-junctions.

Finally, the suppression of the PE due to $e$--$e$ interactions is
expected to be present in other hybrid SN structures as well. In
particular, corrections due to Coulomb blocking would be expected to
be larger when the dimension of the normal metal is smaller than
two. We expect that the main features of the analysis remain valid:
the suppression of the PE with increasing disorder, and its
disappearance at a critical value.
   
Summarizing, we have studied the interplay between the PE and $e$--$e$
interactions in disordered SN-systems. Focusing on an
analysis of the TDoS, it was shown how disorder amplifies the
destructive effect of interactions on the PE. Above a certain
disorder concentration, the suppression is complete in the sense that
the characteristic PE-induced minigap structure in the TDoS
disappears. It was argued that the phenomenon should be of concrete
relevance for the tunnel-spectroscopic observation of the PE in disordered
mesoscopic SN junctions.

We are grateful to A.~V.~Andreev, B.~I.~Halperin, F.~W.~J.~Hekking,
A.~Kamenev and L.~S.~ Levitov for illuminating discussions.  YO is
thankful for the support by the Rothschild fund. The work was
supported by the NSF under grants no.\ DMR 94-16910, DMR 96-30064, and
DMR 97-14725.


\begin{thebibliography}{10}

\vspace{-0.5cm}

\bibitem{NS:Bauernschmitt94} R. Bauernschmitt et al., Phys. Rev. B
   {\bf 49},  4076  (1994).
 
\bibitem{NS:Golubov95} A.~A. Golubov and M. yu. Kuprianov, Pis'ma
Zh. Eksp. Teor. Fiz. {\bf 61}, 830 (1995), [JETP Lett, {\bf 61} 851
(1995)].
 
\bibitem{NS:Beenakker95} C.~W.~J. Beenakker, in {\em Mesoscopic
Quantum Physics}, edited by E. Akkermans et al. (North Holland,
Amsterdam, 1995).
 
\bibitem{NS:Melsen96} J.~A. Melsen et al., Europhys.  Lett. {\bf 35},
7 (1996).
 
\bibitem{NS:Belzig96} W. Belzig et al., Phys. Rev. B {\bf 54}, 9443
(1996).
 
\bibitem{NS:Shechter97} M. Schechter et al., cond-mat/9709248
(unpublished).
 
\bibitem{NS:Zhou98} F. Zhou et al., J. low Temp. Phys. {\bf 110}, 841
(1998).
 
\bibitem{NS:Gueron96} S. Gu\'eron et~al., Phys. Rev. Lett. {\bf 77},
3025 (1996).
  
\bibitem{DS:Altshuler79} B.~L. Altshuler and A.~G. Aronov,
Sol. Stat. Com. {\bf 30}, 115 (1979).
 
\bibitem{DS:Finkelstein90} A.~M. Finkel'stein, in {\em Soviet
Scientific Review}, edited by I.~M.  Khalatnikov (Harwood Academic
Publisher GmbH, Moscow, 1990), Vol.~14.
 
\bibitem{DS:Levitov96} L.~S. Levitov and A.~V. Shytov, Pis'
Zh. Eksp. Teor. Fiz. {\bf 66}, 200 (1997), [JETP Lett, {\bf{66}}, 214
(1997)].
 
\bibitem{NS:Huck97} A.~Huck, F.~W.~J.~Hekking, and B.~Kramer,
Europhys. Lett. {\bf 41}, 201 (1998).

\bibitem{NS:Nazarov96} Y.~V.~Nazarov and T.~H.~Stoof,
Phys. Rev. Lett. {\bf 76}, 823 (1996).

\bibitem{NS:EffectiveDisorder98} We note that the effective disorder
depends not only on the intrinsic material parameters but also on
geometrical factors, e.g. the film thickness, in the slab geometry
discussed in the text.

\bibitem{DSC:Kravtsov91} V.~E. Kravtsov and T.~R. Oppermann,
Phys. Rev. B {\bf 43}, 10865 (1991).
 
\bibitem{NS:Altland98} A. Altland et al., Pis'ma Zh. Eksp. Teor
Fiz. {\bf 67}, 21 (1998), [JETP Lett. {\bf 67}, 22 (1998)].
 
\bibitem{NS:Eilenberger68} G. Eilenberger, Z. Phys. B {\bf 214}, 195
(1968).
 
\bibitem{NS:Usadel70} K.~D. Usadel, Phys. Rev. Lett. {\bf 25}, 507
(1970).
 
\bibitem{NS:footnote98} Note that a more detailed derivation of the
effective action together with a discussion of the symmetry properties
of the matrix fields $Q$ can be found in
Ref.~\protect{\cite{DS:Finkelstein90}}.
 
\bibitem{NS:Kuprianov88} M.~Y. Kuprianov and V.~F. Lukichev,
Zh. Eksp. Teor. Fiz. {\bf 94}, 139 (1988), [Sov. Phys. JETP {\bf 67},
1163 (1988) ].
 
\bibitem{SC:Morel62} P. Morel and P.~W. Anderson, Phys. Rev {\bf 125},
1263 (1962).
 
\bibitem{NS:Phase98} Note that phase fluctuations of the {\it bulk}
  order parameter are inhibited by the Josephson relation and hence do
  not significantly affect the PE.

\bibitem{NS:Likharev79} K.~K. Likharev, Rev. Mod. Phys. {\bf 51}, 101
(1979).
 
\bibitem{DSC:Kamenev98remark} Conceptually, the procedure here is similar to the one used
 in Ref~\protect{\cite{DSC:Kamenev98}}.
 
\bibitem{NS:highenergy98} We note that high momenta ($q \gtrsim 1/d$)
and energies ($\varepsilon\gtrsim E_d$) generate additional
non-singular contributions to $\delta\nu$ which scale linearly in $t$.
 
\bibitem{NS:inlog98} When we use the approximated RG procedure to
estimate the first order correction to the TDoS, we get an expression
that reproduces Eq.~(\protect{\ref{eq:nuI}}) for $\varepsilon \gg E_{\rm
g}$ and has a slightly smaller in-log coefficient for $\varepsilon \sim
E_{\rm g}$.
 
\bibitem{DSC:Kamenev98} A. Kamenev and A. Andreev cond-mat/9810191.
 
\bibitem{DS:Kopietz98} P. Kopietz, Phys. Rev. Lett. {\bf 81}, 2120 (1998).

\bibitem{NS:Schon90} For a general discussion of the charge and
  phase representations of Josephson junction physics, see,
  e.g. G. Sch\"on and A.~D. Zaikin, Phys, Rep. {\bf 98}, 237 (1990).
 
\end{thebibliography}
 \end{document}